\documentclass[preprint,preprintnumbers,amsmath,amssymb]{revtex4}
\usepackage[abs]{overpic}
\usepackage{graphicx,subfigure}
\preprint{HUPD1504}
\begin{document}
\def\nn{\nonumber}
\def\beq{\begin{equation}}
\def\eeq{\end{equation}}
\def\bei{\begin{itemize}}
\def\eei{\end{itemize}}
\def\bea{\begin{eqnarray}}
\def\eea{\end{eqnarray}}
\def\ynu{y_{\nu}}
\def\nub{{\bar{\nu}}}
\def\Hp{{H^+}}
\def\ep{{e^+}}
\def\em{{e^-}}
\def\ydu{y_{\triangle}}
\def\ynut{{y_{\nu}}^T}
\def\ynuv{y_{\nu}\frac{v}{\sqrt{2}}}
\def\ynuvt{{\ynut}\frac{v}{\sqrt{2}}}
\def\s{\partial \hspace{-.47em}/}
\def\ad{\overleftrightarrow{\partial}}
\def\ss{s \hspace{-.47em}/}
\def\pp{p \hspace{-.47em}/}
\def\pdx{\frac{\partial}{\partial x}}
\def\bos{\boldsymbol}
\title{Study of anomalous tau lepton decay using chiral Lagrangian with vector mesons}
\author{Takuya Morozumi, Hiroyuki Umeeda}
\address{ 
        Graduate School of Science, Hiroshima University,
Core of Research for the Energetic Universe,Hiroshima University, Higashi-Hiroshima 739-8526, Japan}
\author{Daiji Kimura}
\address{General Education, Ube National College of Technology, Ube
, Yamaguchi 755-8555, Japan}
\begin{abstract}
An intrinsic parity violating hadronic tau lepton decay is investigated.  $\tau \to \pi \pi \eta \nu$ is the process in which the dominant
contribution to the amplitude is due to the intrinsic parity violation.
To predict the hadronic invariant mass spectra and to compare them with  experimental data,  we extend the chiral Lagrangian
with vector mesons so that it incorporates the intrinsic parity violating terms
and $\phi$ and $\eta^\prime$ mesons.
The coefficients of the intrinsic parity violating terms will be determined 
by fitting the branching fractions for  $V^I \rightarrow P \gamma$,
$V^I \rightarrow 3 P$ and $P \to V^I \gamma $ where $V^I$ denotes vector mesons $1^-$ and $P$ denotes pseudo-scalar 
mesons $0^-$.
\end{abstract}
\maketitle
\section{Introduction}
Hadronic tau decays are important because the precise predictions for the 
process lead to the understanding QCD of energy range from $1$ (GeV) up to $2$ (GeV).
In this energy region, there are many resonances and the description of the
chiral perturbation including only SU(3) octet pseudo-scalar mesons is not 
sufficient. Since there are many resonances, 
perturbative QCD is not applicable. Though there are also many resonance models with which the hadronic form factors
are computed, the theoretical framework which can be systematically improved
has been waited for. Incorporating light SU(3) octet 
vector mesons, a framework which enables us to compute the quantum corrections of pseudo Nambu-Goldstone bosons in a systematic way, has been    
proposed \cite{Kimura:2014wsa}. In the previous work \cite{Kimura:2014wsa}, $\tau \to K \pi \nu$ decay is studied and
the vector and scalar form factors are computed. As for the vector mesons,
$K^\ast$ meson contributes to the form factors. 
The aim of our study is to extend the Lagrangian including intrinsic parity 
violation.  
In the hadronic tau decay including more than two pseudo-scalars as final states of hadrons, the intrinsic parity violation is very important
\cite{Pich:1987qq}. 
For instance, in the hadronic tau decays with final states of odd number of 
pseudo-scalars plus a neutrino,
the vector current contribution is intrinsic parity violating 
 while in the process with the final states with even number of pseudo-scalars plus a neutrino, the axial current contribution is intrinsic parity violating. 
\section{$\tau^- \to \pi^- \pi^0 \eta \nu$}
In the decay $\tau^- \to \pi^- \pi^0 \eta \nu$, the intrinsic parity violating contribution is dominant contribution, because intrinsic parity conserving contribution arises as axial current contribution and it is suppressed due to G parity. The matrix elements of vector form factor  
$\langle \pi^0 \pi^- \eta|\bar{d} \gamma_\mu u| 0 \rangle$ is intrinsic parity violating.
Wess Zumino term \cite{Wess:1971yu} which represents chiral anomaly
contributes to the matrix element. However, the other intrinsic parity violating terms which include the vector mesons also contribute to the process. 
\begin{figure}
\begin{center}
\begin{tabular}{|c|c|} \hline
$V \to \pi^- \pi^0 \eta$ & $\rho^- \to \pi^0 \pi^- \eta$  \\ \hline
\includegraphics[width=4cm]{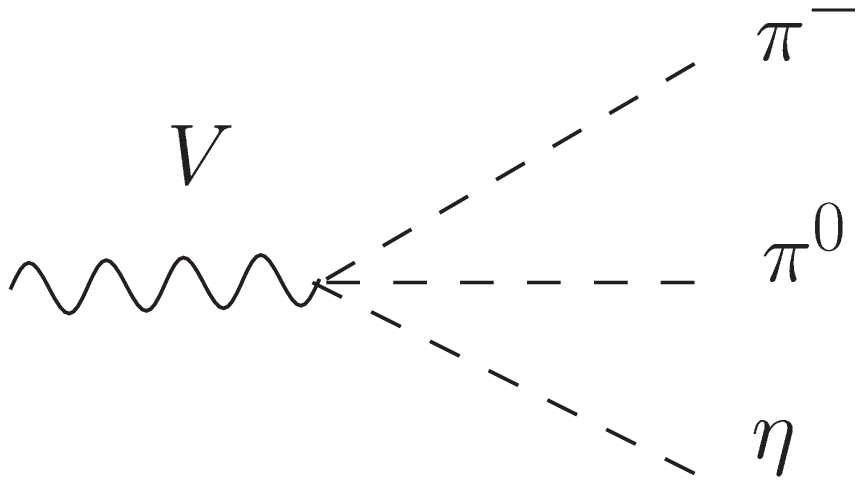} &\includegraphics[width=4cm]{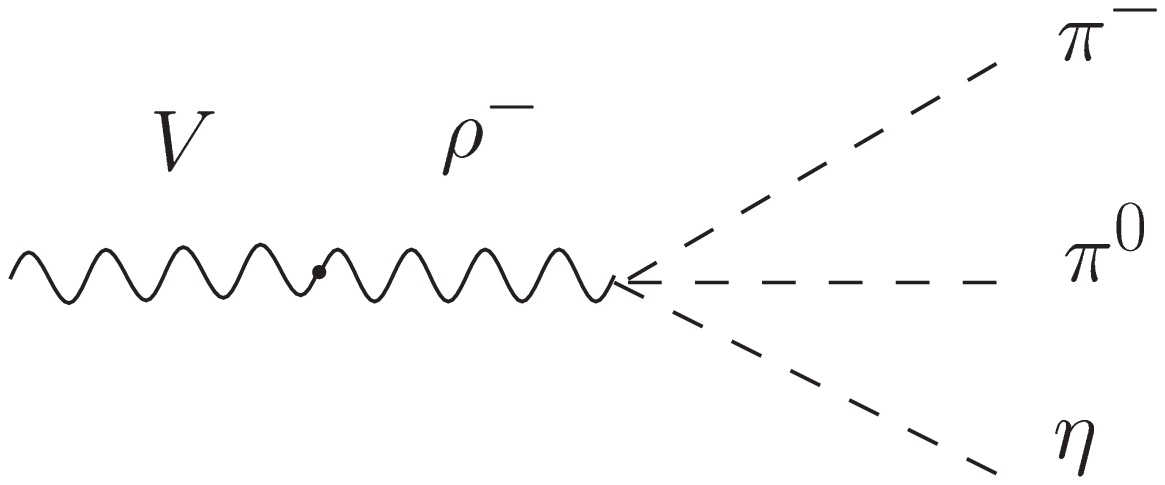} 
\\ \hline
$V \to \rho^- \eta $& $V \to \rho^- \pi^0$ \\ \hline
\includegraphics[width=4cm]{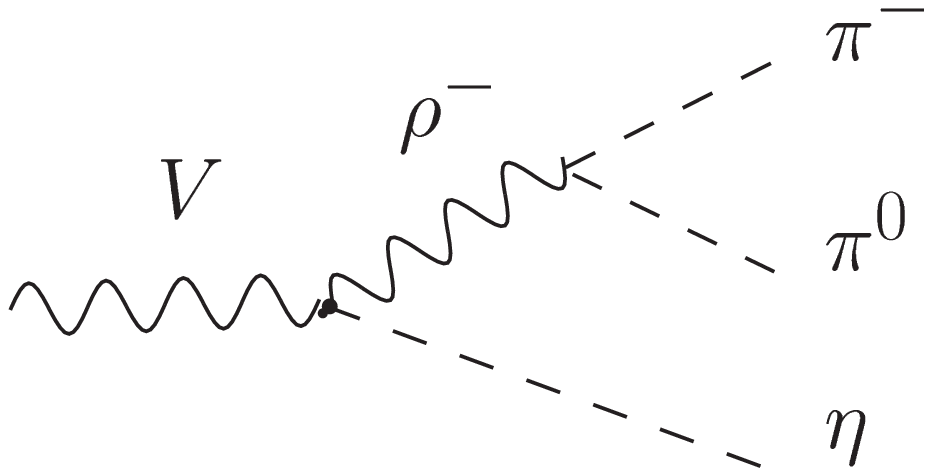} &
\includegraphics[width=4cm]{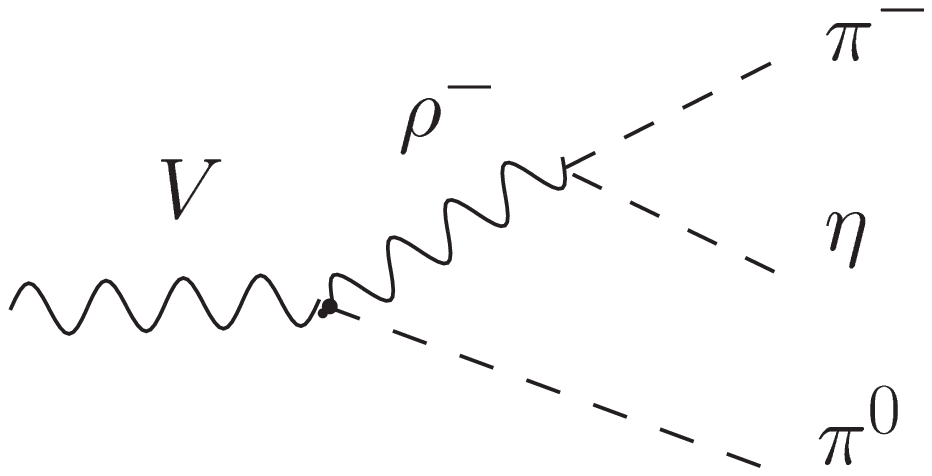} \\ \hline
$\rho^- \to \rho^- \eta $ & $\rho^- \to \rho^- \pi^0$ \\ \hline
\includegraphics[width=4cm]{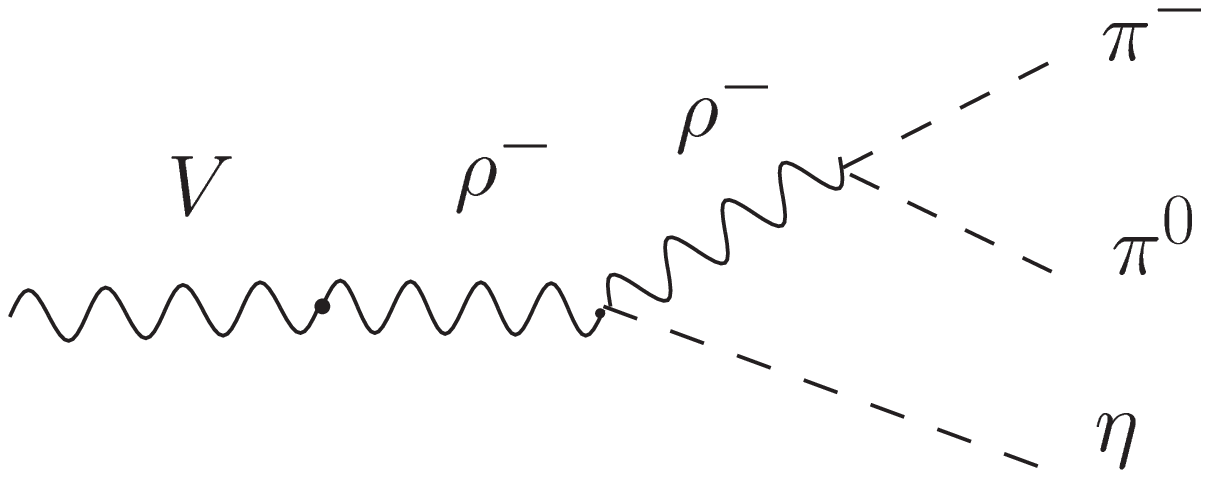} &
\includegraphics[width=4cm]{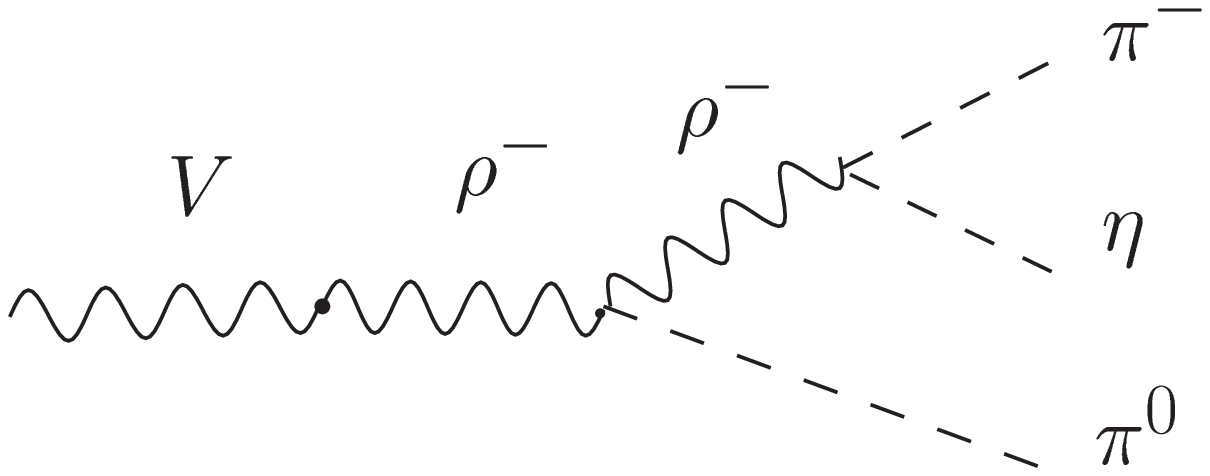} \\ \hline
\end{tabular}
\end{center}
\caption{The Feynman diagrams which contribute to the vector form factor of $\tau \to \pi^- \pi^0 \eta \nu$. V denote the external vector current part of $V-A$
charged current,
$V_\mu =\bar{d} \gamma_\mu u $.} 
\label{fig1}
\end{figure}
As one can see from the Feynman diagrams shown in Fig.1,
there are  many intrinsic parity violating vertices which 
contribute to the decay.   They are $V \rightarrow \pi^- \pi^0
\eta$, $V \to \rho \eta$ and $V \to \rho \pi$ where we denote $V$
as weak gauge boson. 
These vertices are related to $\gamma^* \to 3 P$, $V^I \to P \gamma$, 
where we denote $V^I $ as 
vector mesons  and  $P$ as pseudo-scalars.
One also has vertices like
$ \rho^- \to \rho^- \eta, \rho^- \to \rho^- \pi^0$ 
$ \rho^- \to \pi^- \pi^0 \eta $ and they are related to the processes like 
$V^I  \to V^J P$
and $V^I \to 3 P$.
\section{Chiral Lagrangian with vector mesons with intrinsic parity violation}
Our aim is to extend the chiral Lagrangian and include the intrinsic parity
violating terms.  The intrinsic parity violating terms besides Wess Zumino term
can be written as follows.  
\bea
\mathcal{L}_1&=&i\epsilon^{\mu\nu\rho\sigma}\mathrm{Tr}[\alpha_{L\mu}
\alpha_{L\nu}\alpha_{L\rho}\alpha_{R\sigma}-(R\leftrightarrow L)],
\label{eq2-2}\\
\mathcal{L}_2&=&i\epsilon^{\mu\nu\rho\sigma}\mathrm{Tr}[\alpha_{L\mu}
\alpha_{R\nu}\alpha_{L\rho}\alpha_{R\sigma}],\label{eq2-3}\\
\mathcal{L}_3&=&-\frac{1}
{2}\epsilon^{\mu\nu\rho\sigma}\mathrm{Tr}[F_{V\mu\nu}
\{ \alpha_{L\rho}\alpha_{R\sigma}-(R\leftrightarrow L)\}],
\label{eq2-4}\\
\mathcal{L}_4
&=&\epsilon^{\mu\nu\rho\sigma}\mathrm{Tr}[
(\hat{F}_L+\hat{F}_R)\{ \alpha_{L\rho}, \alpha_{R\sigma}\}],
\label{eq2-4pp}\\
\mathcal{L}_5&=&
\epsilon^{\mu\nu\rho\sigma}F^0_{V\mu\nu}
\mathrm{Tr}[\alpha_{L\rho}\alpha_{R\sigma}-(R\leftrightarrow L)],
\label{eq2-4p}\\
\mathcal{L}_6&=&\frac{\eta_0}{f}
\epsilon^{\mu\nu\rho\sigma}
\mathrm{Tr}F_{V\mu\nu}F_{V\rho\sigma},
\label{eq2-4p2}\\
\mathcal{L}_7&=&\frac{\eta_0}{f}
\epsilon^{\mu\nu\rho\sigma}
F^0_{V\mu\nu}F^0_{V\rho\sigma}.
\label{eq2-4p3}
\eea
In Eq.(\ref{eq2-2})-Eq.(\ref{eq2-4p3}), $\alpha_L$ and $\alpha_R$ include pseudo-scalar octet.
$F_{L \mu \nu}$ and $F_{R \mu \nu}$ are field strength for gauge bosons. 
$F_{V \mu \nu}$ is a field strength of SU(3) octet vector mesons.
$F^0_{\mu \nu}$ denotes the field strength of  SU(3) singlet vector meson $\phi^0$.  Neutral components of octet and singlet vector mesons  form $\rho$, $\phi$ and $\omega$ mesons through 
their mixing. The Lagrangian of intrinsic parity violating terms is given as,
\bea 
{\cal L}=\sum_{i=1}^7 c_{i IPV} {\cal L}_{i}.
\eea
$\mathcal{L}_{1,2,3}$ given in Eqs.(\ref{eq2-2}-\ref{eq2-4})
are introduced in Refs.\cite{Fujiwara:1984mp}
\cite{Bando:1987br}.
We newly consider the operator
in Eqs.(\ref{eq2-4p}-\ref{eq2-4p3}).
In Eqs.(\ref{eq2-2}-\ref{eq2-4p3}), we also required that
the operators should be Hermite.
The Lagrangian is invariant under chiral SU(3) $\times$ SU(3). 
The coefficients can be constrained using $V^I  \rightarrow P 
 \gamma$, $ P \to \gamma \gamma$ and $V^I \to 3 P$.
 Once the coefficients are fixed, one can apply it to the form factor 
calculation of $\tau$ decay. Alternatively, one can use the experimental 
distributions for the decay \cite{Inami:2008ar} to determine the coefficients.
\section{The vector meson propagator and a consistent treatment of  
vector mesons and $\gamma$ mixing in 
$V^I \to P^i \gamma$ amplitude}
As shown in Fig.1,  $\rho^-$
meson contributes to the process.  In the distribution with respect to invariant mass such as $m_{\pi^- \pi^0}$ and $m_{\pi^- \eta}$,
$\rho^-$ should show up as the resonance.
In our framework, the $\rho^-$ meson propagator
is obtained by calculating one-loop corrections to the self energy of the 
vector mesons due to pseudo-scalars.  
In the process $V^I \to P^i \gamma$ in Fig.2,
the left panel does not include the vector meson propagator. 
In the diagram of the right panel, the neutral vector meson propagator is included. The neutral vector mesons correspond to $\rho^0$, $\omega$ and $\phi$ 
and they mix with photon in the final states. Precision of the calculation requires 
us to use one -loop corrected vector meson propagator as in the case of the vector form factor of tau decay. The naive use of the one loop corrected propagator in $V^I \to P^i \gamma$ decay leads to the result which depends on 
the propagators of intermediate vector mesons denoted by $V^J$ in 
Fig.2,
\bea
D_{V^J}(q_\gamma^2=0),
\eea
where $q_\gamma$ is a four momentum of the photon.
$q_\gamma^2=0$ implies the vector mesons $V^J$ mix with on-shell photon $(q_\gamma^2=0)$.
To be consistent, one must use the one-loop 
corrected $\gamma$ and vector mesons' mixing.
By using it, one can show the amplitude does not depend on the propagator of 
the intermediate vector meson at all.
\begin{figure}
\begin{center}
\begin{tabular}{cc} 
\includegraphics[width=4cm]{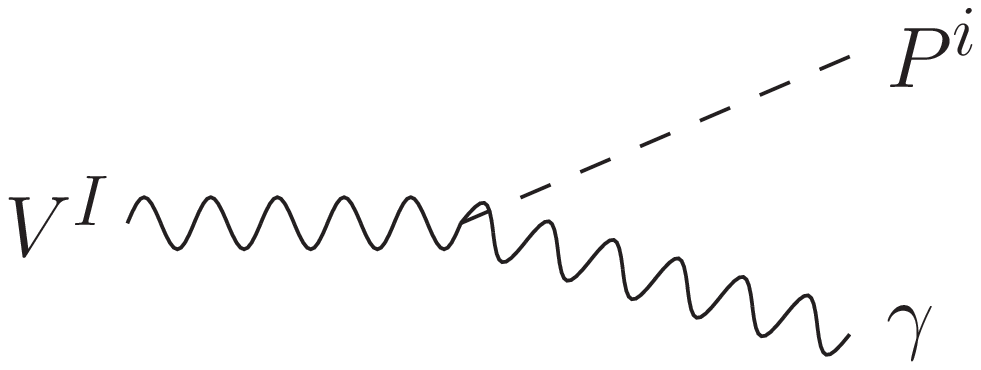} &
\includegraphics[width=4cm]{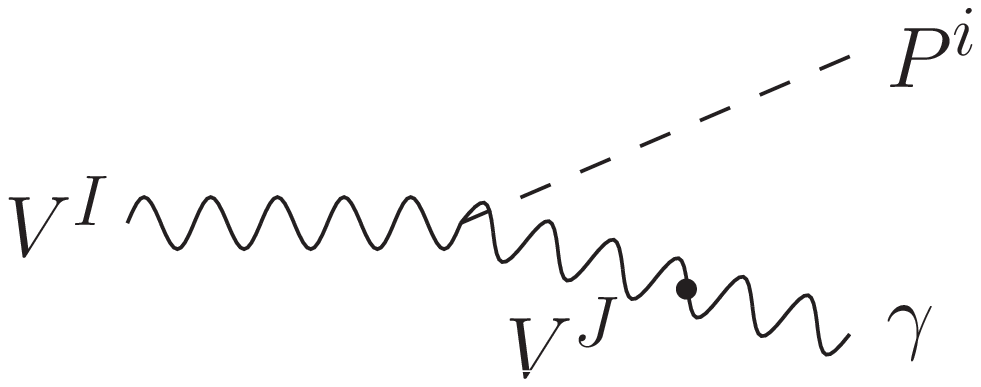} \\
\end{tabular}
\caption{The Feynman diagrams for $V^I \to P^i \gamma$ decay.
The diagram of the right includes intermediate vector meson propagator.}
\end{center}
\label{fig2}
\end{figure}

\end{document}